\documentclass[a4paper, reqno, 11pt]{amsart}
\makeatletter

\@addtoreset{equation}{section}
\makeatother
\usepackage{setspace}
\usepackage{xcolor}
\usepackage{braket}
\usepackage[T1]{fontenc}
\usepackage{amsmath}
\usepackage{amssymb}
\usepackage{latexsym}
\usepackage{amsthm}
\usepackage{textcomp}
\usepackage{hyperref}
\usepackage{here}
\newtheorem{Thm}{Theorem}[section]

\newtheorem{Prop}[Thm]{Proposition}

\theoremstyle{definition}

\newtheorem{Rem}[Thm]{Remark}

\begin{document}

\title{Metrization of powers of the Jensen-Shannon divergence}
\author{Kazuki Okamura}
\address{Department of Mathematics, Faculty of Science, Shizuoka University, 836, Ohya, Suruga-ku, Shizuoka, 422-8529, JAPAN.}
\email{okamura.kazuki@shizuoka.ac.jp}
\maketitle

\begin{abstract}
Metrization of statistical divergences is valuable in both theoretical and practical aspects.  
One approach to obtaining metrics associated with divergences is to consider their fractional powers. 
Motivated by this idea, 
Os{\'an}, Bussandri, and Lamberti (2018) studied the metrization of fractional powers of the Jensen-Shannon divergence between multinomial distributions and posed an open problem.
In this short note, we provide an affirmative answer to their conjecture. 
Moreover, our method is also applicable to fractional powers of $f$-divergences between Cauchy distributions.
\end{abstract}

\section{Introduction}\label{sec:intro}

Dissimilarity between probability distributions is a fundamental topic in probability, statistics, and related fields such as machine learning, and has been extensively studied (\cite{Rachev2013}). 
Statistical divergences serve as canonical measures of dissimilarity. 
One of the most widely used divergences is the Kullback-Leibler divergence (KLD), also known as relative entropy. 
It has numerous theoretical and practical applications. 
In particular, it naturally appears as the rate function in Sanov's theorem in large deviation theory, describing the exponential decay rate of rare events. 
In the framework of information geometry, the KLD generalizes the squared Euclidean distance, and for exponential families, it satisfies a Pythagorean theorem. 
However, the square root of the KLD is generally not a metric: it can be asymmetric and violate the triangle inequality. 
Another commonly used divergence is the total variation distance (TVD). Unlike the KLD, the TVD is a bounded metric. 
However, the TVD between two distributions which are singular to each other always equals $2$. 
Furthermore, closed-form expressions are often difficult to obtain, and one typically must rely on numerical approximations. 

The Jensen-Shannon divergence (JSD), defined via the KLD, is also referred to as the information radius or total divergence from the average. 
The JSD is always well-defined, symmetric, and bounded (\cite{Lin1991}). 
It has found applications across numerous research disciplines and admits both statistical and information-theoretic interpretations. 
In statistical inference, the JSD provides both lower and upper bounds on the Bayes error probability, while in information theory, it is related to mutual information (\cite{Grosse2002}). 
Various generalizations and related notions of the JSD have been proposed (\cite{Nielsen2019j,Osan2022}).

From a theoretical standpoint, metric spaces are one of the most foundational frameworks in mathematics. 
Metrizing divergences is also significant in practical applications, especially in the design of efficient algorithms in computational geometry (\cite{Berg2000}). 
For instance, the triangle inequality can accelerate proximity queries (\cite{Yianilos1993}) and $k$-means clustering (\cite{Elkan2003}). 
In general, symmetric divergences are not metrics, so it is natural to consider fractional powers (moments) of these divergences to obtain associated metric structures. 
Sufficient conditions for fractional powers of Csisz\'ar's $f$-divergences to form metrics are given in \cite{Kafka1991,Osterreicher2003}. 
It is well-known that the square root of the JSD satisfies the triangle inequality (\cite{Endres2003,Vajda2009,Acharyya2013}). 
This, along with the TVD, constitutes a canonical statistical metric distance. 

Os{\'a}n, Bussandri, and Lamberti  \cite{Osan2018} considered the JSD as a special case of a Csisz\'ar divergence and provided a sufficient condition for the power of the JSD between multinomial distributions to define a metric. 
In \cite[Conjecture 1]{Osan2018}, they conjecture that the $p$-th power of the JSD between multinomial distributions is {\it not} a metric for $p > 1/2$. 
The square root of the JSD admits an isometric embedding into Hilbert spaces (\cite{Fuglede2004}). 
Since in general $p$-th powers of distances on Hilbert spaces are {\it not} distances when $p > 1$, this indirectly supports the conjecture. 
However, the embedding is far from being surjective, so this fact cannot be directly used in the proof. 
To the best of our knowledge, this problem remains open. 

One aim of this paper is to prove  \cite[Conjecture 1]{Osan2018}. 
Our approach is elementary and self-contained, differing significantly from \cite{Osan2018}.  
While our method is somewhat similar to the proof of \cite[Theorem 28]{Nielsen2022f},  
we do not utilize the metric transformation introduced there. 
Furthermore, we present an alternative proof of \cite[Proposition 1]{Osan2018}, which is considerably simpler than the original.

Our elementary method also applies to the Cauchy distribution, a canonical example of a heavy-tailed distribution. 
For Cauchy distributions, $f$-divergences are always symmetric (\cite{Nielsen2022f,Verdu2023}), which motivates the question of whether powers of $f$-divergences form metrics for general convex functions $f$. 
We prove that the $p$-th power of the $f$-divergence between Cauchy distributions fails to be a metric for $p > 1/2$, for a broad class of differentiable convex functions f on $(0, \infty)$, including those corresponding to the KLD and the JSD, but excluding the TVD. 
Our proof relies on an expression of $f$-divergences given by Verd\'u \cite{Verdu2023}.

We include a short note in Appendix \ref{sec:sqrt-JSD} on the fact that the square root of the JSD satisfies the triangle inequality. 
This implies that the JSD defines a regular semi-metric, meaning that its local properties are similar to those of a metric. 
See \cite{ChrzaszczJachymskiTurobos2018}  for further details.

\section{Framework}\label{sec:framework}

Let $X$ be a set with a sigma-algebra and $\mu$ be a positive measure on $X$. 
For the discrete and continuous distributions, $\mu$ is usually taken as the counting measure and the Lebesgue measure, respectively. 
Let $P$ and $Q$ be two probability measures on $X$ with density functions $p$ and $q$ with respect to $\mu$, respectively. 
The Kullback-Leibler divergence between $P$ and $Q$ is defined by 
\[ D_{KL}(P : Q) :=  \int_X \log\left(\frac{p(x)}{q(x)}\right) p(x) \mu(dx).  \]
The Jensen-Shannon divergence  between $P$ and $Q$ is defined by 
\[ D_{JS}(P:Q) := \frac{1}{2} \left(D_{KL}\left(P : \frac{P+Q}{2}\right) + D_{KL}\left(Q : \frac{P+Q}{2}\right)\right). \]
There are generalizations of this divergence. 
For example, \cite{Nielsen2019j} replaces $(P + Q)/2$ with a quasi-arithmetic mean. 

We can define them by using the Radon-Nikodym derivative. 
The Kullback-Leibler divergence is asymmetric in general, but the Jensen-Shannon divergence is always symmetric. 
We also remark that $P$ and $Q$ are both absolutely continuous with respect to $(P+Q)/2$, so the Jensen-Shannon divergence is always defined. 
If $P$ is not absolutely continuous with respect to $Q$, then, $D_{KL}(P : Q) = +\infty$. 
These are canonical examples of $f$-divergences. 

We let the entropy be 
\[ H(P) := \int_X -p(x) \log p(x) \mu(dx). \]
Then, 
\[ D_{JS}(P:Q) = H\left(\frac{P+Q}{2}\right) - \frac{H(P) + H(Q)}{2}. \]

We now recall the definition of a metric. 
Let $S$ be a non-empty set. 
We call a function $d : S \times S \to [0, \infty)$ a distance function if it satisfies the following three conditions: \\
(1) $d(x,y) = 0$ if and only if $x=y$. \\
(2) (symmetry) $d(x,y) = d(y,x)$ for $x, y \in S$. \\
(3) (triangle inequality) $d(x,z) \le d(x,y) + d(y,z)$ for $x, y, z \in S$.\\
For such $d$, we call a pair $(S, d)$ a metric space. 
This is a fundamental notion in geometry.

\section{Metrization of Jensen-Shannon divergences between the multinomial distributions}\label{sec:JSD-MN}

Throughout this section, we set $\log = \log_2$, so that $0 \le D_{JS}(P:Q) \le 1$, and, 
$D_{JS}(P:Q)$ is smaller than the total variation distance.  
The natural-log version differs only by a constant factor of $\ln 2$. 

We  let $X = \{1,2, \cdots, n\}$ and $\mu$ be the counting measure on $X$. 
For $n \ge 2$, let $\mathcal{P}_n := \{(p_i)_i : \sum_i p_i = 1, p_i > 0\}$ and $\overline{\mathcal{P}_n}  := \{(p_i)_i : \sum_i p_i = 1, p_i \ge 0\}$.   
For $P = (p_i)_{i=1}^{n}$ and $Q = (q_i)_{i=1}^{n}$ in $\mathcal{P}_n$, 
\[ D_{KL}(P : Q) = \sum_{i=1}^{n} p_i \log_2 \left(\frac{p_i}{q_i}\right), \]
and 
\[ D_{JS}(P:Q) = \sum_{i=1}^{n} - \frac{p_i}{2} \log_2 \left(\frac{p_i + q_i}{2p_i}\right) - \frac{q_i}{2} \log_2 \left(\frac{p_i + q_i}{2q_i}\right).  \]

Our main result is 
\begin{Thm}\label{thm:multi}
Let $\alpha > 1/2$. 
Then, $D_{JS}(P:Q)^{\alpha}$ is not a metric on $\mathcal{P}_n$. 
\end{Thm}

\begin{proof}
We first deal with the case that $n=2$. 
Let $P_t := (t, 1-t)$, $0 \le t \le 1$. 
For $t \in [0, 1/2)$, let 
$f(t) := D_{JS}(P_{1/2 - t}:P_{1/2 + t})$ 
and 
$g(t) := D_{JS}(P_{1/2 - t}:P_{1/2}) = D_{JS}(P_{1/2}:P_{1/2 + t}).$ 
Let 
$F(t) := f(t)^{\alpha} - 2g(t)^{\alpha}$. 
It suffices to show that $F(t) > 0$ for some $t$. 
Since $F(0) = 0$, by the mean-value theorem, 
it suffices to show that $F^{\prime}(t) > 0$ for every  $t$  sufficiently close to $0$. 
Since 
$\displaystyle F^{\prime}(t) = \alpha (f^{\prime}(t)f(t)^{\alpha-1} - 2g^{\prime}(t)g(t)^{\alpha-1})$, 
it suffices to show that 
\begin{equation}\label{eq:wts-1}
\left(\frac{g(t)}{f(t)}\right)^{1-\alpha} > 2 \frac{g^{\prime}(t)}{f^{\prime}(t)}, 
\end{equation}
 for every  $t$ sufficiently close to $0$. 
 
We see that $\displaystyle f(t) = 1 - H(P_{1/2 + t})$ and $\displaystyle g(t) = H(P_{(1+t)/2}) - \frac{H(P_{1/2 + t}) + 1}{2}.$
Hence, 
$$f^{\prime}(t) = - \frac{d}{dt} H(P_{1/2 + t}) \textup{ and } g^{\prime}(t) = \frac{d}{dt} H(P_{(1+ t)/2}) - \frac{1}{2} \frac{d}{dt}H(P_{1/2 + t}).$$
Since $H(P_s) = -s \log_2 s - (1-s) \log_2 (1-s), \, 0 \le s \le 1$, 
we see that 
$\displaystyle \frac{d}{dt} H(P_{(1+ t)/2}) = - \frac{1}{2} \log \left(\frac{1+t}{1-t}\right)$ and $\displaystyle \frac{d}{dt} H(P_{1/2 + t}) = - \log \left(\frac{1+2t}{1-2t}\right).$ 
Hence, 
\[ 2 \frac{g^{\prime}(t)}{f^{\prime}(t)} = 1 - 2 \frac{\frac{d}{dt} H(P_{(1+ t)/2})}{\frac{d}{dt} H(P_{1/2 + t})} = 1 - \frac{\log \frac{1+t}{1-t}}{\log\frac{1+2t}{1-2t}}. \]
Since $\displaystyle \lim_{s \to 0} \frac{\log\left(\frac{1+s}{1-s}\right)}{2s} = 1,$
we see that $\displaystyle \lim_{t \to 0} 2 \frac{g^{\prime}(t)}{f^{\prime}(t)} = \frac{1}{2}.$ 

We recall that $f(0) = g(0) = 0$. 
Then, by l'Hospital's theorem, 
$$\lim_{t \to 0} \frac{g(t)}{f(t)} = \lim_{t \to 0} \frac{g^{\prime}(t)}{f^{\prime}(t)} = \frac{1}{4}.$$ 
Hence, $\displaystyle \lim_{t \to 0} \left(\frac{g(t)}{f(t)}\right)^{1-\alpha} = \left(\frac{1}{4}\right)^{1-\alpha}.$
Since $\alpha > 1/2$,
$\displaystyle \left(\frac{1}{4}\right)^{1-\alpha} > \frac{1}{2}$. 
Thus we have Eq.~\eqref{eq:wts-1}. 
The proof of Theorem 1 is completed for $n=2$.  

We now deal with the case of $n \ge 3$. 
We can naturally embed $\overline{\mathcal{P}_2}$ into $\overline{\mathcal{P}_n}$ by a map $(p_1, p_2) \mapsto (p_1, p_2, 0, \cdots, 0)$. 
Since $P \mapsto H(P)$ is continuous with respect to $P$ on $\overline{\mathcal{P}_n}$, 
we can find $P_1, P_2$ and $P_3$ in $\mathcal{P}_n$ such that 
$\displaystyle D_{JS}(P_1:P_3)^{\alpha} > D_{JS}(P_1:P_2)^{\alpha} + D_{JS}(P_2:P_3)^{\alpha}.$ 
The proof of Theorem \ref{thm:multi} is completed for $n \ge 3$.  
 \end{proof}

\begin{Rem}
In general, $x^{\beta} + y^{\beta} \le (x+y)^{\beta}$ for $x, y \ge 0$ and $\beta \ge 1$, 
and, if $x^{\beta} + y^{\beta} = (x+y)^{\beta}$, then, $\beta = 1$ or $xy=0$. 
Hence, if a function $d: S \times S  \to [0, \infty)$ is {\it not} a metric on  a set $S$, then,  $d^{\beta}(x,y)$ is {\it not} a metric $S$. 
Since it is known that $D_{JS}(P:Q)^{1/2}$ is a metric, 
this gives an alternative proof of \cite[Proposition 1]{Osan2018}. 
which is much easier than the proof given in it. 
\end{Rem}

\section{Metrization of $f$-divergences between the Cauchy distributions}\label{sec:JSD-Cauchy}

For $\mu \in \mathbb{R}$ and $\sigma > 0$, the density function of the univariate Cauchy distribution is given by 
$\displaystyle p_{\mu, \sigma}(x) := \frac{\sigma}{\pi} \frac{1}{(x -\mu)^2 + \sigma^2}, \ x \in \mathbb{R}.$ 
For a continuous function $f$ on $(0,\infty)$, the $f$-divergence is defined by 
$$D_f (p_{\mu_1, \sigma_1} : p_{\mu_2, \sigma_2}) := \int_{\mathbb{R}} f\left(\frac{p_{\mu_2, \sigma_2} (x)}{p_{\mu_1, \sigma_1} (x)}\right) p_{\mu_1, \sigma_1} (x) dx.$$

The following result is crucial in our proof. 
\begin{Thm}[{\cite[Eq.~(189) in Theorem 10]{Verdu2023}}]\label{thm:Verdu}
Let $f$ be a continuous function on $(0, \infty)$. Then, 
\[ D_f (p_{\mu_1, \sigma_1} : p_{\mu_2, \sigma_2}) = \int_0^{\pi} f\left(\frac{1}{\zeta + \sqrt{\zeta^2 -1} \cos\theta}\right) \frac{d\theta}{\pi}, \]
where $\displaystyle \zeta := 1 + \frac{(\mu_2 - \mu_1)^2 + (\sigma_2 - \sigma_1)^2}{2\sigma_1 \sigma_2}$. 
\end{Thm}

In particular, every $f$-divergence is a function of $\zeta$. 
This quantity is also known as maximal invariant with respect to an action of the special linear group $SL(2, \mathbb R)$ to the upper-half plane $\mathbb{H} := \{\mu+\sigma i : \mu \in \mathbb{R}, \sigma > 0\}$ with complex parameter, considered by McCullagh \cite{McCullagh1996}. 
For example, we obtain the JSD if we let 
\[ f(u) = f_{JS}(u) := \frac{1}{2} \left(u \log \frac{2u}{1+u} - \log \frac{1+u}{2} \right).\]

\begin{Thm}
Let $f$ be a convex function on $(0, \infty)$ such that $f(1) = 0$, $f$ is in $C^2$ class on an open neighborhood of $1$, and $f^{\prime\prime}(1) > 0$.  
Let $\alpha > 1/2$. 
Then, $D_{f}(p_{0, \sigma_1}: p_{0, \sigma_2})^{\alpha}$ is not a metric on $(0, \infty)$. 
\end{Thm}

This result is applicable to a large class of $f$-divergences including the KLD and the JSD. 
However, the regularity assumption for $f$ is crucial. 
Obviously, the conclusion fails for the TVD, which is obtained by $f(u) = f_{TV}(u) := |u-1|/2$.

\begin{proof}
We will show that 
$$D_{f}(p_{0, \sigma_1}: p_{0, \sigma_2})^{\alpha} + D_{f}(p_{0, \sigma_2}: p_{0, \sigma_3})^{\alpha} < D_{f}(p_{0, \sigma_1}: p_{0, \sigma_3})^{\alpha}$$
where 
$(\sigma_1, \sigma_2, \sigma_3) = (e^{-t}, 1, e^{t})$ for sufficiently small $t > 0$. 
For $t > 0$,  let
\[ h(t) := \int_0^{\pi} f\left(\frac{1}{\cosh(t) + \sinh(t) \cos\theta}\right) \frac{d\theta}{\pi}. \]
Then, by Theorem \ref{thm:Verdu}, 
$h(t) = D_{f}(p_{0, \sigma_1}: p_{0, \sigma_2}) = D_{f}(p_{0, \sigma_2}: p_{0, \sigma_3})$ and $h(2t) = D_{f}(p_{0, \sigma_1}: p_{0, \sigma_3})$. 
Hence, it suffices to show that $2h(t)^{\alpha} < h(2t)^{\alpha}$ for some $t > 0$. 
We remark that 
\begin{equation}\label{eq:lim1}
\lim_{t \to +0} \cosh(t) + \sinh(t) \cos\theta = 1
\end{equation} 
and 
\begin{equation}\label{eq:lim2}
\lim_{t \to +0} \sinh(t) + \cosh(t) \cos\theta = \cos\theta \in [-1,1]. 
\end{equation}

Under the assumption of $f$, 
we can exchange the derivative wit respect to  $t$ and the integral wit respect to  $\theta$, 
so we obtain that there exists a sufficiently small $\delta_0 > 0$ such that for every $0 < t < \delta_0$, 
 \[ h^{\prime}(t) =  \int_0^{\pi} -\frac{\sinh(t) + \cosh(t) \cos\theta}{(\cosh(t) + \sinh(t) \cos\theta)^2} f^{\prime}\left(\frac{1}{\cosh(t) + \sinh(t) \cos\theta}\right) \frac{d\theta}{\pi}, \]
 and, 
 \[ h^{\prime\prime}(t) =  \int_0^{\pi} \frac{(\sinh(t) + \cosh(t) \cos\theta)^2}{(\cosh(t) + \sinh(t) \cos\theta)^4} f^{\prime\prime}\left(\frac{1}{\cosh(t) + \sinh(t) \cos\theta}\right) \frac{d\theta}{\pi} \]
 \[ +  \int_0^{\pi} \frac{2(\sinh(t) + \cosh(t) \cos\theta)^2 - (\cosh(t) + \sinh(t) \cos\theta)^2}{(\cosh(t) + \sinh(t) \cos\theta)^3} f^{\prime}\left(\frac{1}{\cosh(t) + \sinh(t) \cos\theta}\right) \frac{d\theta}{\pi}. \]
We recall that $\displaystyle \int_0^{\pi} \cos\theta d\theta =  \int_0^{\pi} \cos(2\theta) d\theta = 0$ and $\displaystyle \int_0^{\pi} \cos^2 \theta d\theta = \frac{\pi}{2}$. 
By this, \eqref{eq:lim1}, and \eqref{eq:lim2}, we see that 
$$\lim_{t \to +0} h(t) = \lim_{t \to +0} h^{\prime}(t) = 0$$ 
and 
$$\lim_{t \to +0} h^{\prime\prime}(t) = \frac{f^{\prime\prime}(1)}{2} > 0.$$
By l'Hospital's theorem, 
\[ \lim_{t \to +0} \frac{h(2t)}{h(t)} =  \lim_{t \to +0} \frac{2h^{\prime}(2t)}{h^{\prime}(t)} = \lim_{t \to +0} \frac{4h^{\prime\prime}(2t)}{h^{\prime\prime}(t)} = 4. \]
Since $\alpha > 1/2$, we see that $2h(t)^{\alpha} < h(2t)^{\alpha}$ for sufficiently small $t > 0$. 
This completes the proof. 
\end{proof}

\begin{Rem}
(i) In the case of the TVD, 
$\displaystyle \lim_{t \to +0} h^{\prime} (t) = \frac{1}{\pi} > 0$, and hence, by l'Hospital's theorem,  
we have that $\displaystyle \lim_{t \to +0} \frac{h(2t)}{h(t)} = 2$. \\
(ii) In \cite[Theorem 10]{Verdu2023}, it is assumed that $f$ is convex and right-continuous at $0$. 
However, for every $(\mu_1, \sigma_1)$ and $(\mu_2, \sigma_2)$, 
$$0 < \inf_{x \in \mathbb{R}} \frac{p_{\mu_2, \sigma_2}(x)}{p_{\mu_1, \sigma_1}(x)} \le \sup_{x \in \mathbb{R}} \frac{p_{\mu_2, \sigma_2}(x)}{p_{\mu_1, \sigma_1}(x)} < +\infty,$$
so we do not need to assume that $f$ is defined at $0$. 
This property does not hold for normal distributions. 
\end{Rem}

\appendix
\section{On the square root of Jensen-Shannon divergence}\label{sec:sqrt-JSD} 

Fuglede and Tops{\o}e \cite{Fuglede2004} stated that the square root of the JSD is a metric on the space of probability measures over a given measure space. 
Acharyya, Banerjee, and Boley \cite{Acharyya2013} provided a proof of this result. 
However, some parts of the arguments of \cite{Fuglede2004,Acharyya2013} are sketchy, and we offer more details here. 
While we follow the overall strategy used in \cite{Fuglede2004,Acharyya2013},  we believe that several components of our approach are more elementary, transparent, and simpler than those in \cite{Acharyya2013}. 
Our arguments make use of the {\it Lambert W function}.

Let $P, Q$ be two probability measures on a measurable space $X$. 
Let $M := (P+Q)/2$. 
Let the Jensen-Shannon divergence between $P$ and $Q$ be  
\[ D_{JS}(P : Q) :=  \frac{1}{2}\left(\int_X \log \frac{dP}{dM} dP + \int_X \log \frac{dQ}{dM} dQ\right).  \]

Let $\phi(z) := z\log z, z \ge 0$. 
Then, this is convex. 
Let 
\[  \psi(x,y) := \sqrt{\frac{\phi(x)+ \phi(y)}{2} - \phi\left(\frac{x+y}{2}\right)}, \ x, y \ge 0.\]  

Let $\lambda$ be a probability measure on $X$ such that $P \ll \lambda$ and $Q \ll \lambda$. 
For ease of notation, we let $f := dP/d\lambda$ and $g := dQ/d\lambda$. 
Then, 
\[ D_{JS}(P : Q) = \int_X \psi(f,g) d\lambda. \]

Let $P, Q, R$ be three probability measures on a measure space $X$. 
Let $\lambda$ be a probability measure on $X$ such that $P \ll \lambda$, $Q \ll \lambda$ and $R \ll \lambda$. 
Let $f := dP/d\lambda$, $g := dQ/d\lambda$ and $h = dR/d\lambda$.  
By the Minkowski inequality, 
in order to show that 
\[ \sqrt{D_{JS}(P : R)} \le \sqrt{D_{JS}(P : Q)} + \sqrt{D_{JS}(Q : R)}, \]
which is equivalent to 
\[  \int_X \psi(f,g) d\lambda \le  \int_X \psi(f,g) d\lambda +  \int_X \psi(f,h) d\lambda, \] 
it suffices to show that 
\begin{Prop}\label{prop:reduction-gen-JSD}
\[ \sqrt{\psi(x,z)} \le \sqrt{\psi(x,y)} + \sqrt{\psi(y,z)}, \ x, y, z \ge 0. \]
\end{Prop}

By Schoenberg's theorem  \cite{Schoenberg1938}, 
in order to show Proposition \ref{prop:reduction-gen-JSD}, it suffices to show that 
\begin{Prop}[{\cite[Lemma 4]{Acharyya2013}}]\label{prop:reduction-gen-JSD-2}
If $k(x,y) := \phi(x+y) = (x+y) \log(x+y)$, then, 
$(x,y) \mapsto \exp(\beta k(x,y))$ is a positive-definite kernel for every $\beta > 0$. 
\end{Prop}

Let $W(x)$ be the inverse function of a $C^{\infty}$ function $z \mapsto z \exp(z)$ on $(-1,\infty)$. 
This is called the Lambert W function, and $W \in C^{\infty}((-1/e,\infty))$. 
By \cite{Kalugin2011},  
$W(\cdot)$ is a Bernstein function. 
Since a map $x \mapsto 1/(1+x)$ is a completely monotone function, 
by \cite[Theorem 3.7 (ii)]{Schilling2010}, 
a map $x \mapsto 1/(1+W(x))$ is also a completely monotone function. 

Hence, by Bernstein's theorem (cf. \cite[Theorem 1.4]{Schilling2010}), 
there exists a unique probability measure $\mu$ on $(0,\infty)$ such that 
\[ \int_0^{\infty} \exp(-tx) \mu(dx) = \frac{1}{1+W(t)}, \ t > 0.  \]

Let $0 < s < 1$. 
Then, by a disintegration formula (cf. \cite[p63]{Zolotarev1986}), 
\[ \int_0^{\infty} x^s \mu(dx) = \frac{s}{\Gamma(1-s)} \int_0^{\infty} t^{-s-1} \left(1- \int_0^{\infty} \exp(-tx) \mu(dx) \right) dt. \]

We see that 
\[ \int \left(1- \int_0^{\infty} \exp(-tx) \mu(dx) \right) dt = \int t^{-s-1} \frac{W(t)}{1+W(t)} dt = -s^{s-1} \Gamma(1-s, sW(t)) + C, \]
where $\Gamma(,)$ is the incomplete Gamma function and $C$ is the integral constant. 

Since $\displaystyle \lim_{t \to +0} W(t) = 0$ and $\displaystyle \lim_{t \to +\infty} W(t) = +\infty$, 
\[ \int_0^{\infty} x^s \mu(dx) = s^s = \exp(\phi(s)), \ 0 < s < 1.\]

We remark that for every $n \ge 1$ and $t > 0$, 
\[ \sup_{x > 0} x^n \exp(-tx) < +\infty. \]
Hence, for each $n \ge 1$, 
\[ \frac{\partial^n}{\partial t^n} \int_0^{\infty} \exp(-tx) \mu(dx)  = \int_0^{\infty} (-x)^n \exp(-tx) \mu(dx), \ t > 0. \]
By the monotone convergence theorem, 
for each $n \ge 1$, 
\[ \int_0^{\infty} x^{2n} \mu(dx) = \lim_{t \to +0} \frac{\partial^{2n}}{\partial t^{2n}} \int_0^{\infty} \exp(-tx) \mu(dx) = \lim_{t \to +0} \frac{\partial^{2n}}{\partial t^{2n}} \frac{1}{1+W(t)}. \]
This limit is finite since $W \in C^{\infty}((-1/e,\infty))$. 
Hence, 
\[ \int_0^{\infty} x^s \mu(dx) < +\infty, \ s > 0. \]
Hence, 
\[ F(z) := \int_0^{\infty} x^z \mu(dx), \ z \in \left\{z \in \mathbb{C} : \textup{Re}(z) > 0 \right\}, \]
is well-defined and holomorphic. 
By the identity theorem for holomorphic functions, 
\[ \int_0^{\infty} x^s \mu(dx) = s^s = \exp(\phi(s)), \ s > 0.\] 

Let $\displaystyle \nu := \mu \circ (\log)^{-1} = \mu \circ \exp$ be a probability measure. 
Then, 
\[ \int_{-\infty}^{\infty} \exp(sy) \nu(dy) = s^s = \exp(\phi(s)), \ s > 0.\] 

Let $\beta > 0$. 
Then, 
\[ \int_{-\infty}^{\infty} \exp(s(\beta y -\log \beta)) \nu(dy) = s^s = \exp(\beta\phi(s)), \ s > 0.\] 

Let $b_1, \cdots, b_n > 0$ and $c_1, \cdots, c_n \in \mathbb{R}$. 
Then, 
\[ \sum_{i,j=1}^{n} c_i c_j \exp(\beta \phi(b_i + b_j)) = \int_{-\infty}^{\infty} \left(\sum_{i=1}^{n} c_i \exp(b_i (\beta y -\log \beta)) \right)^2 \nu(dy) \ge 0. \]
This completes the proof of Proposition \ref{prop:reduction-gen-JSD-2}. 

\begin{Rem}
We see that 
\begin{equation}\label{eq:Cauchy-asym-char} 
\int_{-\infty}^{\infty} \exp(ity) \nu(dy) = \exp\left(-\frac{\pi}{2}|t| + i t \log |t|\right), \ t \in \mathbb{R}. 
\end{equation}
Hence, 
$\nu$ is an asymmetric stable distribution with $\alpha = 1$. 
The function $\displaystyle \phi(t) := \int_{-\infty}^{\infty} \exp(ity) \nu(dy)$ is not an entire function, so we cannot apply \cite[Theorem 1]{Ehm2004}. 

The arguments in the proof of \cite[Lemma 4]{Acharyya2013} implicitly assumes that if \eqref{eq:Cauchy-asym-char} holds, 
then, $\displaystyle \int_{-\infty}^{\infty} \exp(sy) \nu(dy) < +\infty$ for every $s > 0$. 
However, the proof is not written in it. 
One easy way to resolve this is to use an integral expression of the density function $g$ of $\nu$ given in \cite[Theorem 2.2.3]{Zolotarev1986} as follows: 
\[ g(x) = \frac{1}{2} \int_{-1}^{1} U(t) \exp(x - \exp(x) U(t)) dt, \ x \in \mathbb{R}, \]
where 
$$U(t) := \frac{\pi}{2} \frac{1-t}{\cos(\pi t/2)} \exp\left(-\frac{\pi}{2} (1-t) \tan(\frac{\pi}{2} t)\right).$$

We see that 
$$\lim_{t \to -1+0} U(t) = \frac{1}{e}, \ \ \lim_{t \to 1-0} U(t) = +\infty,$$ 
and, 
\[ \int_{-1}^{1} U(t) \exp(- \exp(x) U(t)) dt \le \exp(-(n+2)x) \int_{-1}^{1} \frac{(n+2)!}{U(t)^{n+1}} dt, \ n \ge 1. \]
Then, we see that for every $n \ge 1$, 
$$\int_{\mathbb R} \exp(nx) g(x) dx < +\infty. $$
Now we can use the identity theorem as above, 
and obtain that 
$$\int_{\mathbb R} \exp(sx) g(x) dx = \exp(\phi(s)), \ s > 0. $$ 
\end{Rem}

\vspace{1pc}

{\it Acknowledgements} \  
The author wishes to give his thanks to the referee for his or her comments, and to Prof. Frank Nielsen for notifying me of the conjecture by Os{\'a}n, Bussandri, and Lamberti.

\bibliographystyle{unsrt}
\bibliography{metjsd}
\end{document}